\documentclass[prb,aps,superscriptaddress,twocolumn]{revtex4}
\usepackage[dvips]{graphicx}
\usepackage{color}
\usepackage{amssymb}
\usepackage{sidecap}

\def\up{\uparrow}
\def\dw{\downarrow}
\def\ch{\mathrm{ch}}
\def\sp{\mathrm{sp}}
\def\cu{\mathrm{Cu}}
\def\py{\mathrm{Py}}
\def\updw{{\up/\dw}}
\def\ss{\mathrm{ss}} 
\def\is{s} 
\def\side{\mathrm{side}}

\newlength{\tw}
\def\jbox#1{\rotatebox{90}{#1}}
\def\tabI{%
\small
\begin{table*}
\begin{ruledtabular}
\begin{tabular}{lccccccccc}
& \jbox{experiment}  
& \jbox{$AR^\star_{\py/\cu}=0$, $AR_{s,\py/\cu}=$inf} 
& \jbox{$AR_{\ss,\cu}=0.15$} 
& \jbox{$AR_{\ss,\cu,\side}=0.065$}
& \jbox{$AR^\star_{\py/\cu}\approx0$, $AR_{s,\py/\cu}=2.6$}
& \jbox{$AR^\star_{\py/\cu}=0.3$, $AR_{s,\py/\cu}=3.1$}
& \jbox{$AR^\star_{\py/\cu}=1$, $AR_{s,\py/\cu}=3.8$}
& \jbox{$AR^\star_{\py/\cu}=3$, $AR_{s,\py/\cu}=8.2$}
& \jbox{$AR^\star_{\py/\cu}=15$, $AR_{s,\py/\cu}\approx\inf$}
\\ \hline \hline
DNLVS$_\mathrm{half}$, $w_\cu=100$ & 0.7 & 6.1 & 0.77 & 0.74 &
   0.78 & 0.81 & 0.76 & 0.87 & 0.79
\\ \hline
DNLVS$_\mathrm{cross}$, $w_\cu={100}$ & 0.62 & 6.06 & 0.75 & 0.73 &
   0.76 & 0.79 & 0.75 & 0.87 & 0.79
\\ \hline
DNLVS$_\mathrm{half}$, $w_\cu={300}$ & 0.6 & 3.1 & 0.66 & 0.87 &
   0.49 & 0.47 & 0.40 & 0.40 & 0.29
\\ \hline
DNLVS$_\mathrm{cross}$, $w_\cu={300}$  & 0.3 & 1.2 & 0.07 & 0.24 &
  0.08 & 0.11 & 0.14 & 0.21 & 0.25
\\ \hline \hline
3-wires, no middle & 0.25 & 3.72 & 0.034 & 0.032 &
  0.45 & 0.47 & 0.46 & 0.54 & 0.48
\\ \hline
3-wires, Cu middle & 0.18 & 2.53 & 0.025 & 0.033 &
  0.32 & 0.33 & 0.31 & 0.35 & 0.28
\\ \hline
3-wires, Py middle & 0.04 & 0.76 & 0.019 & 0.018 &
  0.037 & 0.052 & 0.075 & 0.16 & 0.33
\\ \hline \hline
DNLVS$_\mathrm{half}$, $w_\cu={250}$ & $\times$ & 0.51 & 2.0 & 0.61 &
  0.29 & 0.29 & 0.25 & 0.27 & 0.25
\\ \hline
DNLVS$_\mathrm{cross}$, $w_\cu={250}$ & 0.4 & 1.3 & 0.20 & 0.33 &
  0.12 & 0.14 & 0.15 & 0.21 & 0.24
\\ \hline
DLVS$_\mathrm{2Py}$, $w_\cu={250}$ & 1 & 3.8 & 2.0 & 2.2 &
  0.50 & 0.50 & 0.46 & 0.51 &0.50
\\ \hline
DLVS$_\mathrm{1Py}$, $w_\cu={250}$ & 0.4 & 2.4 & 1.7 & 1.8 &
  0.34 & 0.32 & 0.28 & 0.29 & 0.26
\end{tabular}
\end{ruledtabular}
\caption{Experimental values of DNLVS (in m$\Omega$) for various
  sample structures compared with values calculated by 3D models
  taking into account different processes decreasing DNLVS. Units of
  resistances are in f$\Omega$m$^2$, units of wire widths in nm. 
  For details see Sec.~\ref{s:nlvs}.}
\label{t:tab1}
\end{table*}
}

\def\figI{%
\begin{figure}[b]
\includegraphics[width=0.3\textwidth]{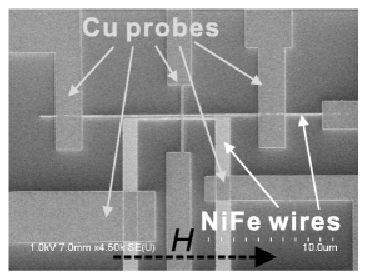}
\caption{%
 SEM image of the fabricated lateral spin-valve device.}
\label{f:elimage}
\end{figure}
}

\def\figII{%
\begin{figure*}
\begin{minipage}{0.7\textwidth}
\begin{tabular}{ccc}
  \includegraphics[scale=0.45]{w2.1}
& \includegraphics[scale=0.45]{w3.1}
& {\raise-0mm\hbox{\includegraphics[scale=0.45]{wl.1}}}
\end{tabular}
\end{minipage}
\begin{minipage}{0.27\textwidth}
\caption{DNLVS for system with (a) 2-wires and (b) with 3-wires. (c)
  Difference of \textit{local} voltage signal (DLVS) for system with 2
  wires.}
\label{f:loc}
\end{minipage}
\end{figure*}
}

\def\figIII{%
\begin{figure*}
\newlength{\wx}
\setlength{\wx}{0.24\textwidth}
\newlength{\wy}
\setlength{\wy}{0.2\textwidth}
\begin{minipage}{0.7\textwidth}
\begin{tabular}{ccc}
& ``half'' & ``cross''
\\
\raisebox{2.5mm}{%
\includegraphics[width=\wy]{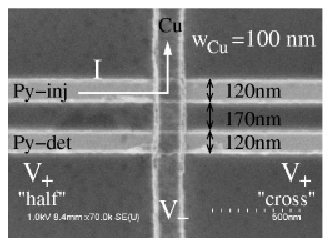}%
\put(-19.5,12.5){\textcolor{white}{(a)}}
}
&
\includegraphics*[width=\wx,angle=0]{100-half}%
\put(-20,3){(c)}
&
 \includegraphics*[width=\wx,angle=0]{100-cross}%
\put(-20,3){(e)}
\\
\raisebox{2.5mm}{%
\includegraphics[width=\wy]{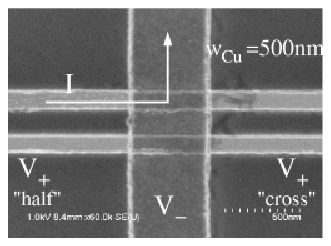}%
\put(-19.5,12.5){\textcolor{white}{(b)}}
}
&
 \includegraphics*[width=\wx,angle=0]{500-half}%
\put(-20,3){(d)}
& 
 \includegraphics*[width=\wx,angle=0]{500-cross}%
\put(-20,3){(f)}
\end{tabular}
\end{minipage}
\begin{minipage}{0.28\textwidth}
\caption{(a)(b) SEM image of a detail of the lateral spin-valve device
  with $w_\cu=100$, 500\,nm, respectively, with sketched current flows
  and ``cross'' and ``half'' detection configuration.  (c-f) NLVS as a
  function of external magnetic field, obtained for ``cross'' and
  ``half'' configuration for $w_\cu=100$, 500\,nm.}
\label{f:loops}
\end{minipage}
\end{figure*}
}

\def\figIV{%
\begin{figure}
  \begin{tabular}{c}
    \includegraphics*[width=0.35\textwidth]{fig-wcuII}
\\
    \includegraphics*[width=0.35\textwidth]{fig-wcu-models1}
  \end{tabular}
\caption{Experimental value of DNLVS as a function of Cu wire
  width $w_\cu$ compared with 1D and 3D models. Experimental data
  (square) compared with 1D and 3D models.  $R^\star$ and $R_s$ are
  interface and surface scattering resistances for Py/Cu interface,
  units in f$\Omega$m$^2$. For detail see Sec.~\ref{s:curr1d}
  and~\ref{s:nlvs}.}
\label{f:wcu}
\end{figure}
}

\def\figV{%
\begin{figure}
  \includegraphics[width=0.3\textwidth]{res.1}%
  \put(-9.8,10.2){%
    \includegraphics[width=0.07\textwidth,angle=0]{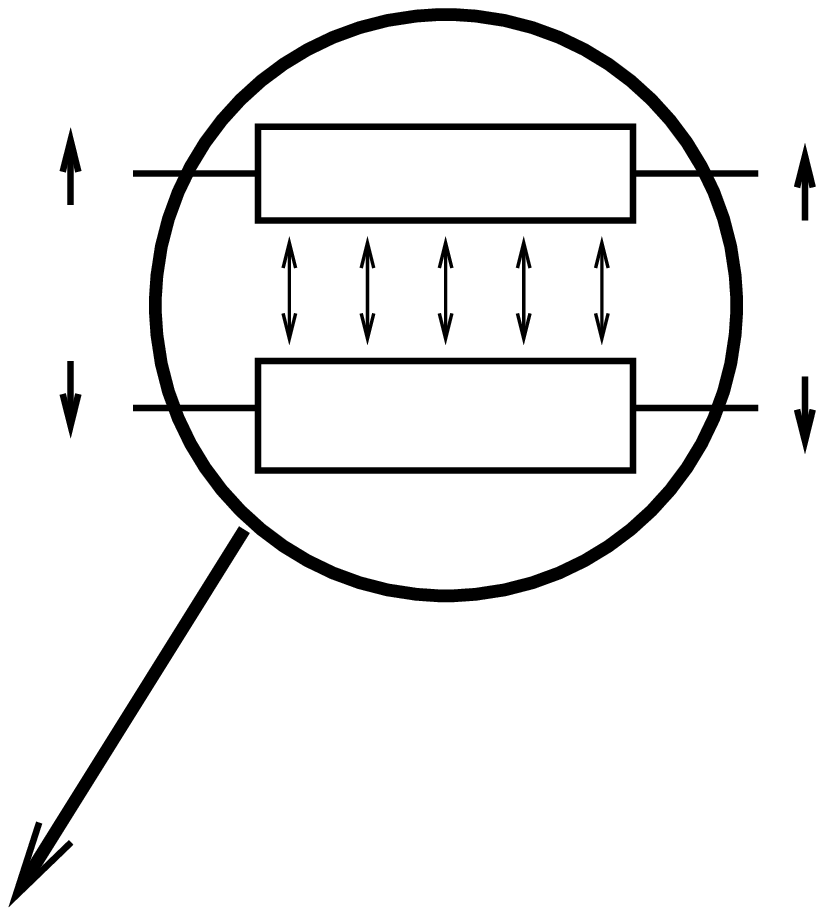}}
  \caption{The sketch of 3D network of spin-dependent-resistance-elements
    (SDRE). Circle inset sketches that SDRE consists of spin-up and
    spin-down resistances and of the shunting resistors between up
    and down channels. Note that each node and wire on the sketch
    represents a ``bus'' containing spin-up and spin-down channels.}
  \label{f:net3d}
\end{figure}
}

\def\figVI{%
\begin{figure}
\includegraphics[scale=0.27]{pycumulap-jr}
\caption{%
  The profile of $j_\sp=j_\up-j_\dw$ through
  Cu/Py(20)/Cu(20)/Py(20)/Cu pillar structure, dimensions in nm,
  calculated for 1D VF model (full line) and compared with our 3D
  calculations with perpendicular-to-interface grid size 1\,nm, 5\,nm,
  10\,nm. Lateral (parallel-to-interface) grid size is 10\,nm.}
\label{f:jr13}
\end{figure}
}

\def\figVII{%
\begin{figure}
  \includegraphics[scale=0.27]{agreeI}
\caption{Dependence of DNLVS on lateral grid size. Details in
  Sec.~\ref{s:plau}.} 
\label{f:gridsize}
\end{figure}
}

\def\figVIII{%
\begin{figure}
\includegraphics[scale=0.7]{cut.1}
\caption{The sketch of the device with indicated cut planes. The yz,
  xz$'$ cuts are taken in the center of Py-injector, Cu wire,
  respectively. The xy cut is located 12.5\,nm from device top, the
  xz cut is located 7.5\,nm from the side of Cu wire.}
\label{f:cut}
\end{figure}
}

\def\figIX{%
\begin{figure*}
\begin{minipage}{0.7\textwidth}
\begin{tabular}{rccrc}
\rotatebox{90}{\hspace*{8mm} $j_\up$}&
\includegraphics[width=0.3\tw]%
  {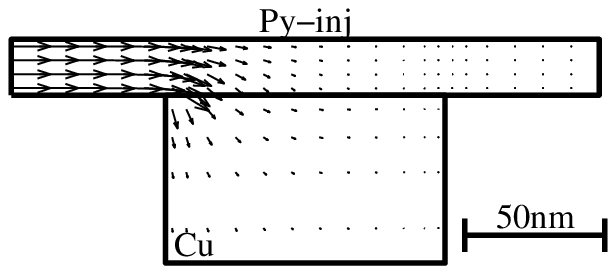}
\put(-30,1){(a)}
&\quad &
\rotatebox{90}{\hspace*{8mm} $j_\ch$}&
\includegraphics[width=0.3\tw]%
  {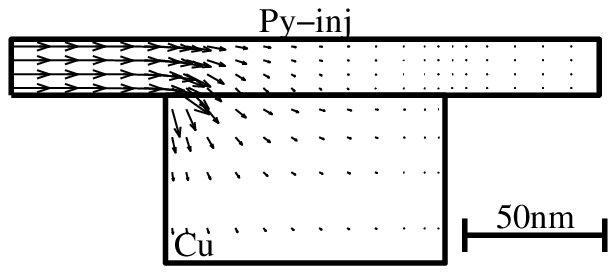}
\put(-30,1){(c)}
\\
\rotatebox{90}{\hspace*{8mm} $j_\dw$}&
\includegraphics[width=0.3\tw]%
  {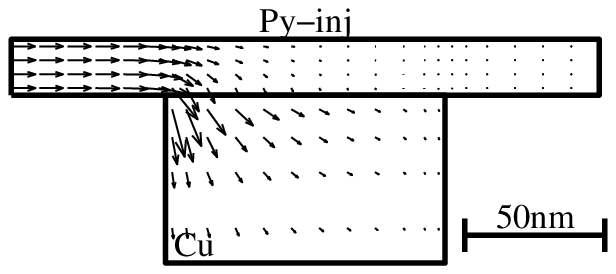}
\put(-30,1){(b)}
&&
\rotatebox{90}{\hspace*{8mm} $j_\sp$}&
\includegraphics[width=0.3\tw]%
  {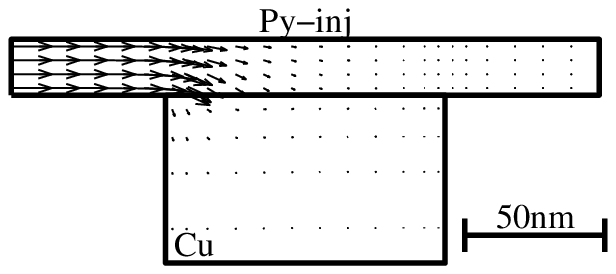}
\put(-30,1){(d)}
\end{tabular}
\end{minipage}
\begin{minipage}{0.27\textwidth}
\caption{The yz cut (defined on Fig.~\ref{f:cut}) of 
  the current density of (a) current polarized up $j_\up$ (b)
  current polarized down $j_\dw$ (c) charge current
  $j_\ch=j_\up+j_\dw$ and (d) spin-polarized current
  $j_\sp=j_\up-j_\dw$ in the device with $w_\cu=100$\,nm with parallel
  magnetization.  The length of arrow is proportional to value of a
  given current;  this scaling is the same for all cuts.}
\label{f:frontinj}
\end{minipage}
\end{figure*}
}

\def\figX{%
\begin{figure}
\includegraphics*[height=0.35\textwidth]{currinj}
\caption{The profiles of $j_\ch$ (open symbols) and $j_\sp$
  (solid symbols) taken at the intersection of Py-injector/Cu
  interface and the yz cut (Fig.~\ref{f:cut}). The profiles were
  calculated for $w_\cu=100$\,nm, for current I=1\,mA, for
  $AR^\star_\mathrm{int}=0$, $AR_{s,\py/\cu}=\inf$ (circle) and for
  $R^\star_\mathrm{int}=1$f$\Omega$m$^2$,
  $AR_{s,\py/\cu}=3.8$\,f$\Omega$m$^2$ (triangles and diamond).
  $j_\sp$ and $j_\ch$ are the same for parallel and anti-parallel
  magnetizations. The symbols denote grid, for which the current
  densities are calculated.}
\label{f:currinj}
\end{figure}
}

\def\figXI{%
\begin{figure*}
\begin{minipage}{0.7\textwidth}
\begin{tabular}{rc}
\rotatebox{90}{\hspace*{12mm} $j_\up$}&
\includegraphics[width=0.6\tw]%
  {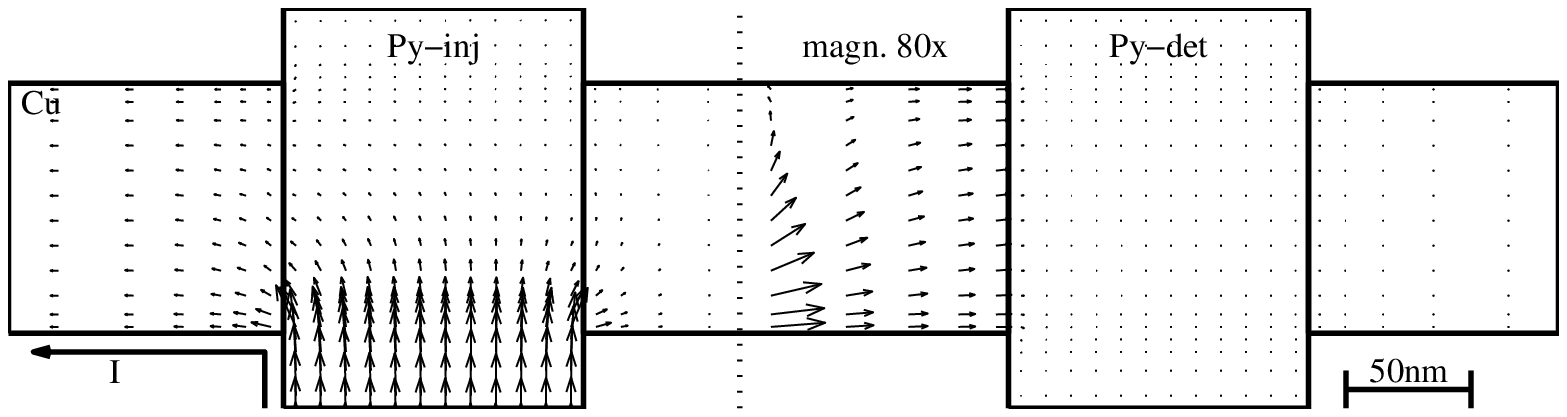}
\put(-60,13){(a)}
\\
\rotatebox{90}{\hspace*{12mm} $j_\dw$}&
\includegraphics[width=0.6\tw]%
  {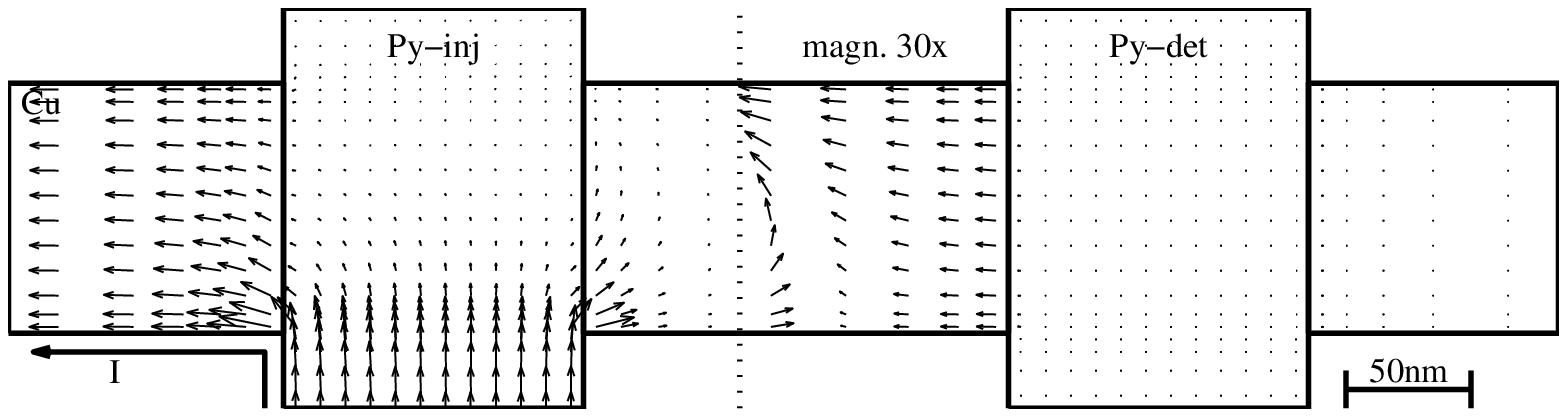}
\put(-60,13){(b)}
\\
\rotatebox{90}{\hspace*{12mm} $j_\ch$}&
\includegraphics[width=0.6\tw]%
  {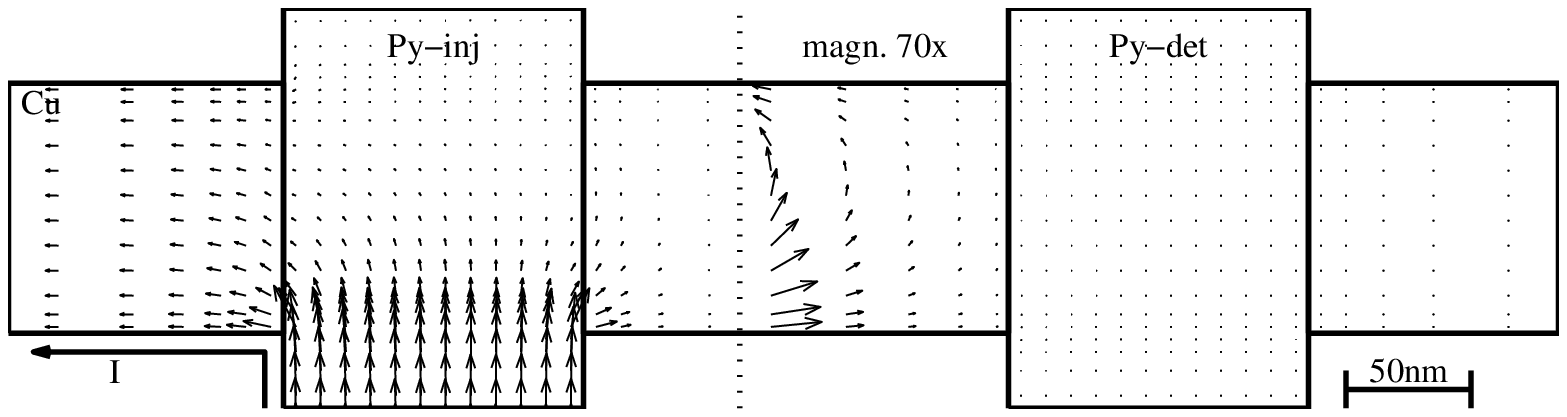}
\put(-60,13){(c)}
\\
\rotatebox{90}{\hspace*{12mm} $j_\sp$}&
\includegraphics[width=0.6\tw]%
  {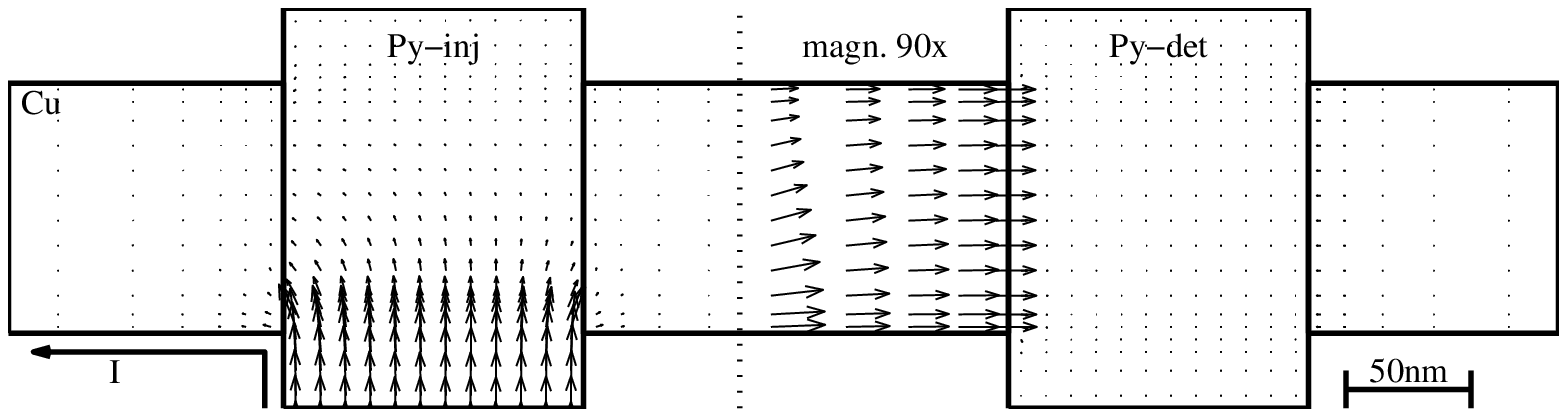}
\put(-60,13){(d)}
\end{tabular}
\end{minipage}
\begin{minipage}{0.27\textwidth}
\caption{The xy cut (defined on Fig.~\ref{f:cut})  of (a)
  $j_\up$ (b) $j_\dw$ (c) $j_\ch$ and (d) $j_\sp$ in the device with
  $w_\cu=100$\,nm, parallel magnetization.  The arrows have the same
  scaling for all cuts, and they are magnified on cut's right sides.}
\label{f:topcu}
\end{minipage}
\end{figure*}
}

\def\figXII{%
\begin{figure}
\begin{tabular}{rc}
\rotatebox{90}{\hspace*{18mm} $j_\ch$}&
\includegraphics[width=0.35\textwidth]%
  {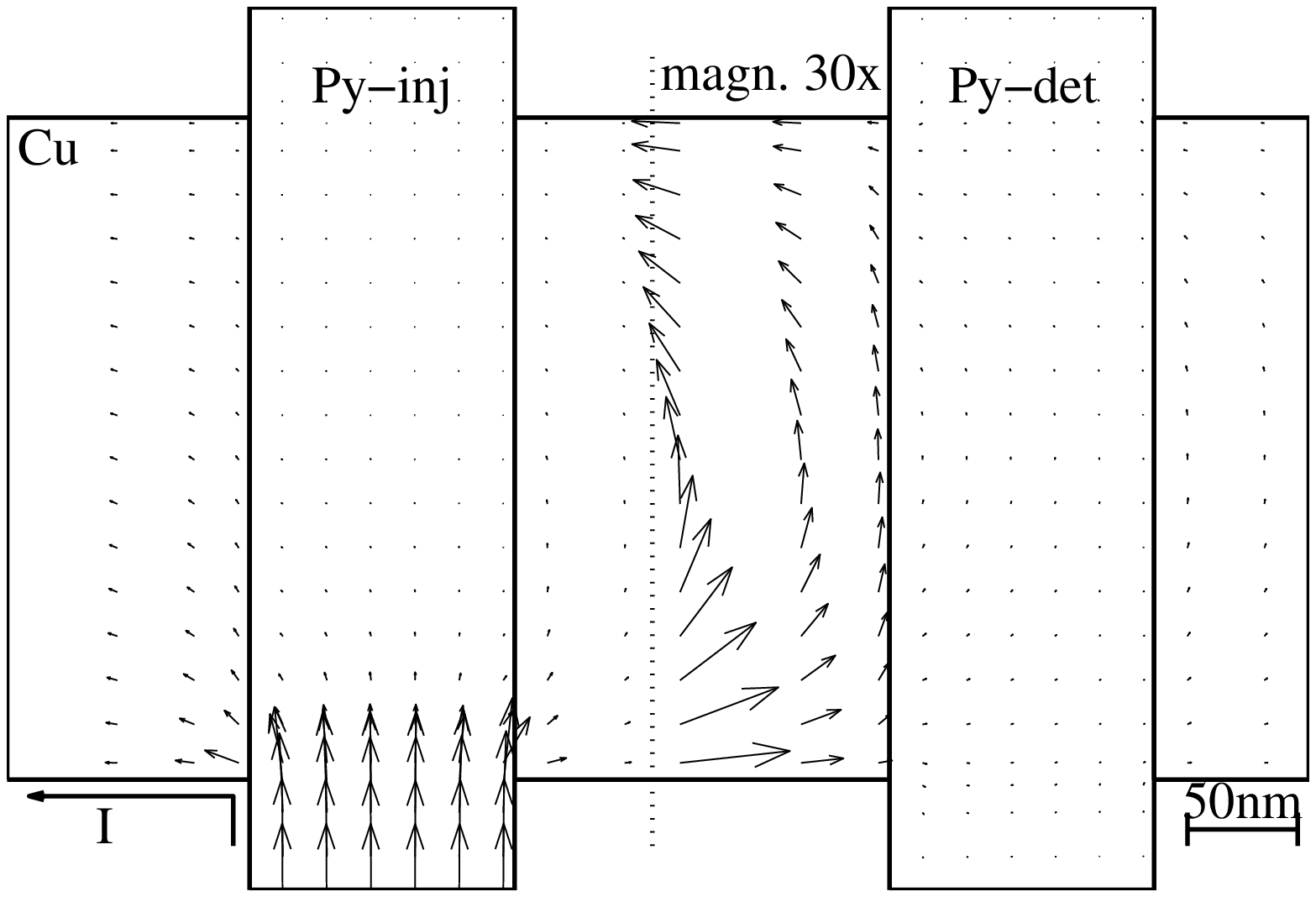}
\put(-35,22){(a)}
\\
\rotatebox{90}{\hspace*{18mm} $j_\sp$}&
\includegraphics[width=0.35\textwidth]%
  {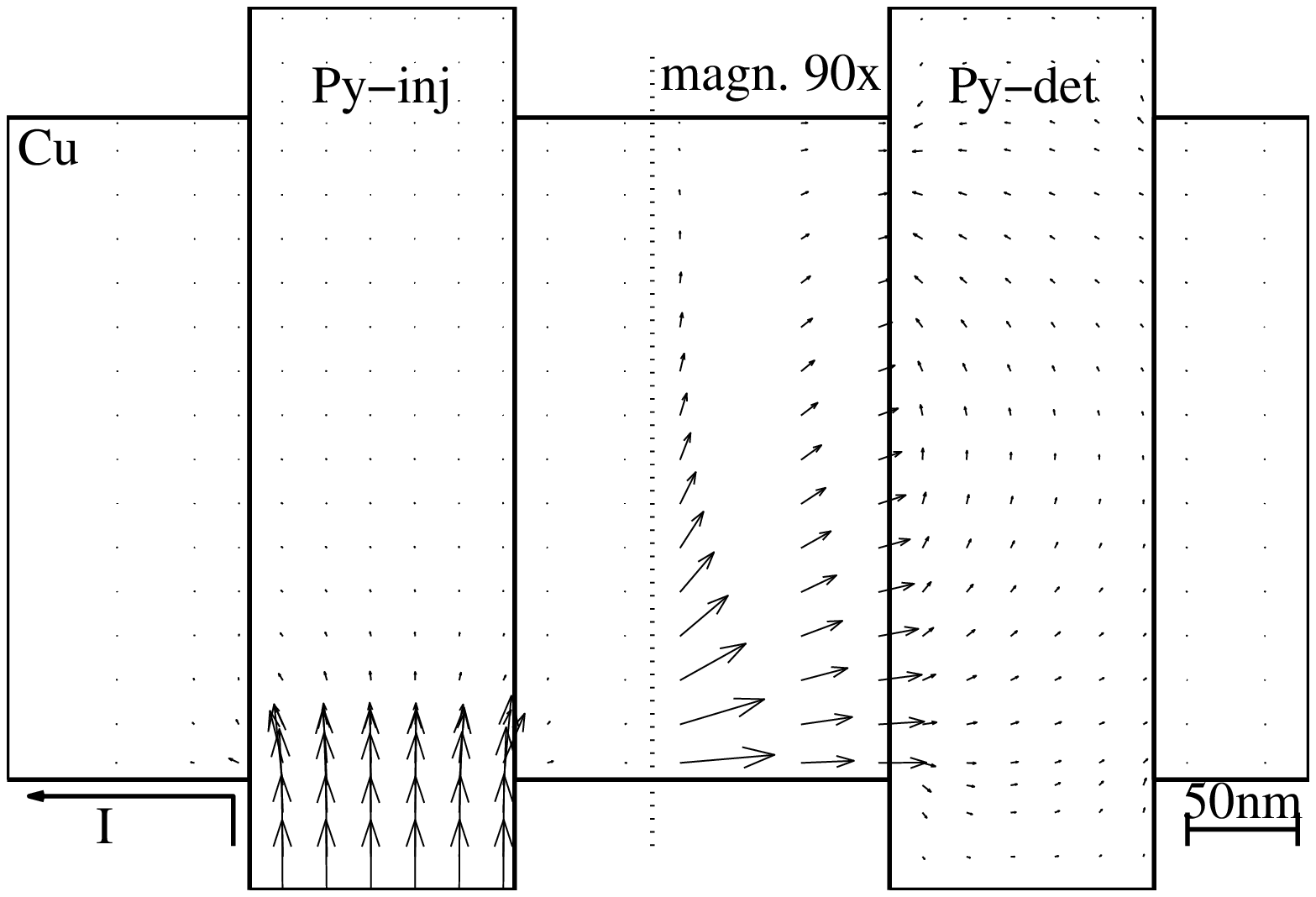}
\put(-35,22){(b)}
\end{tabular}
\caption{The xy cut (defined on Fig.~\ref{f:cut})  of (a) 
  $j_\ch$ and (b) $j_\sp$ in the device with
  $w_\cu=300$\,nm.  The arrows have the same scaling for all cuts, and
  they are magnified on cut's right sides.}
\label{f:wcu300}
\end{figure}
}

\def\figXIII{%
\begin{figure*}
\begin{minipage}{0.7\textwidth}
\begin{tabular}{rc}
\rotatebox{90}{\hspace*{10mm} $j_\up$}&
\includegraphics[width=0.6\tw]%
  {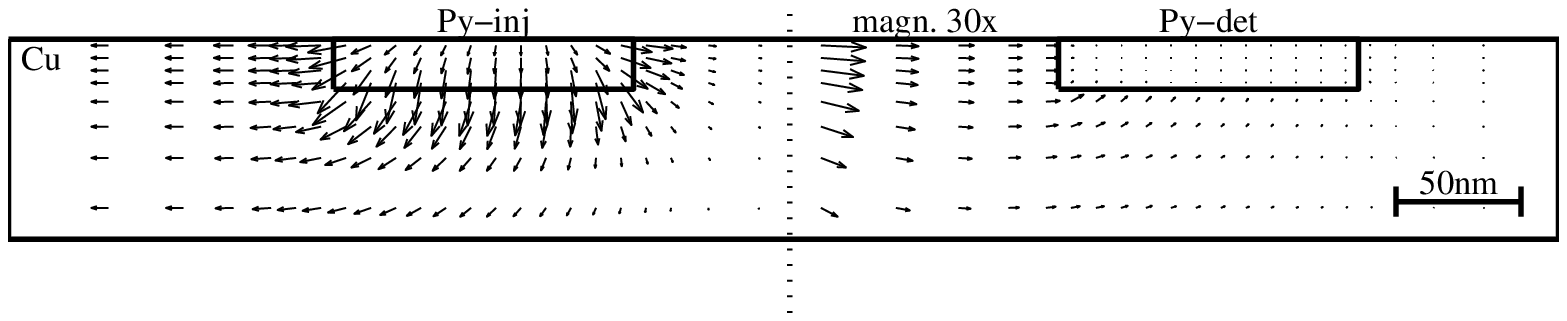}
\put(-59.5,4){(a)}
\\[-5mm]
\rotatebox{90}{\hspace*{10mm} $j_\dw$}&
\includegraphics[width=0.6\tw]%
  {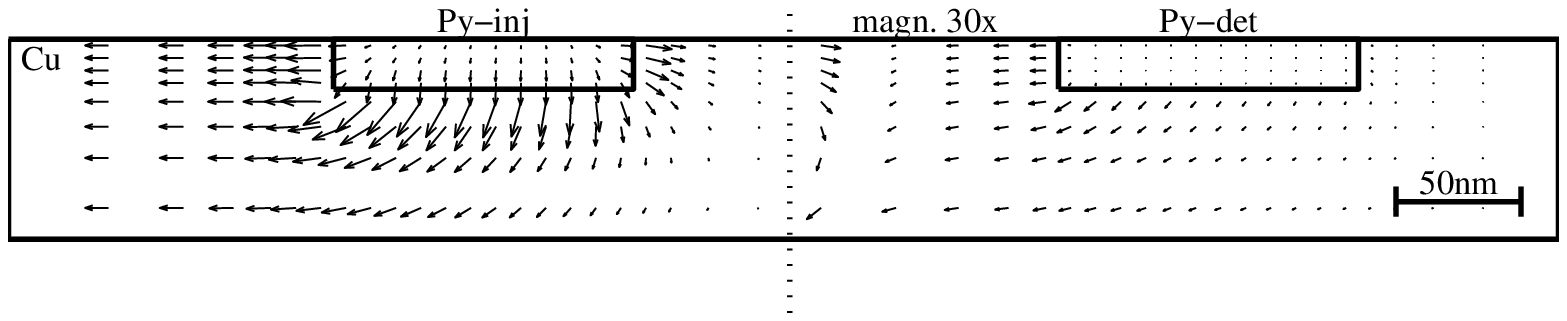}
\put(-59.5,4){(b)}
\\[-5mm]
\rotatebox{90}{\hspace*{10mm} $j_\ch$}&
\includegraphics[width=0.6\tw]%
  {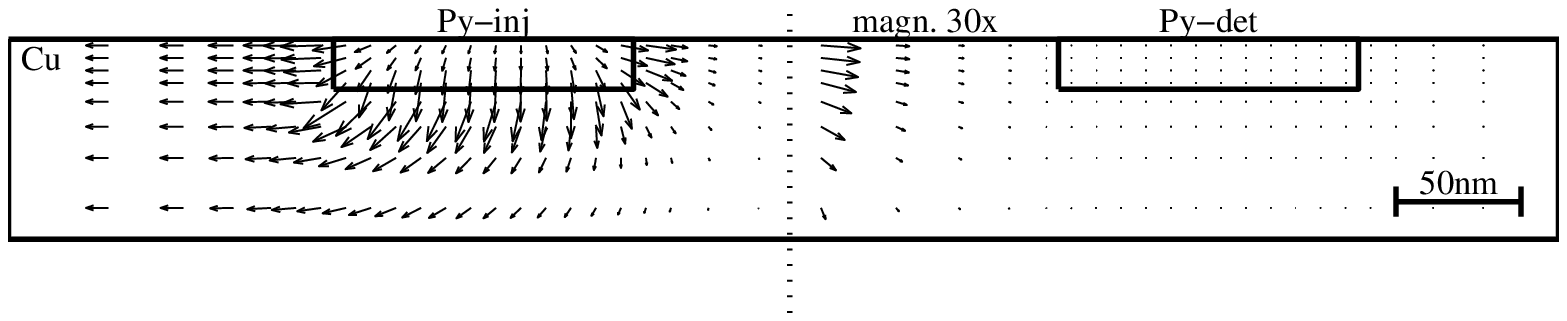}
\put(-59.5,4){(c)}
\\[-5mm]
\rotatebox{90}{\hspace*{10mm} $j_\sp$}&
\includegraphics[width=0.6\tw]%
  {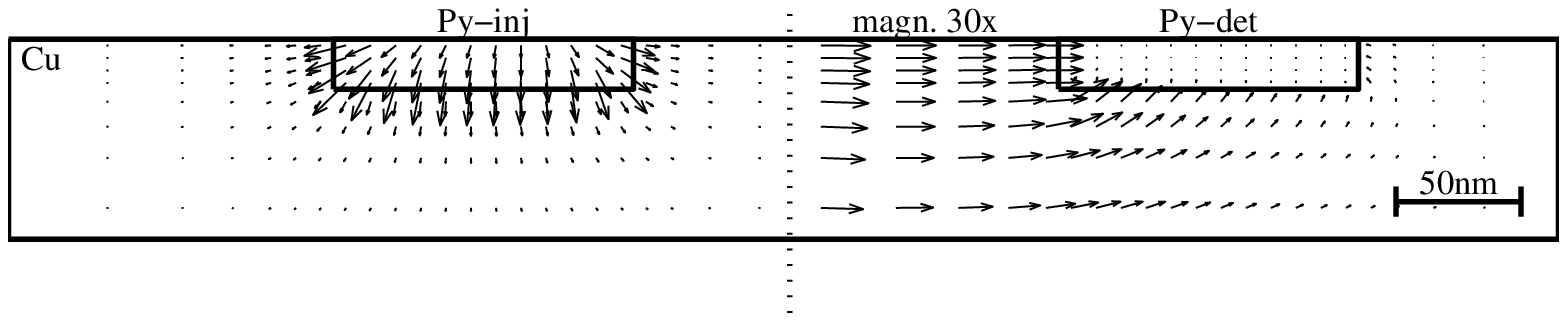}
\put(-59.5,4){(d)}
\end{tabular}
\vspace*{-12mm}\mbox{}
\end{minipage}
\begin{minipage}{0.28\textwidth}
\caption{The xz cut (defined on Fig.~\ref{f:cut}) of (a) 
  $j_\up$ (b) $j_\dw$ (c) $j_\ch$ and (d) $j_\sp$ in the device with
  $w_\cu=100$\,nm, parallel magnetization.  The arrows have the same
  scaling for all cuts, and they are magnified on cut's right sides.}
\label{f:sidecu}
\end{minipage}
\end{figure*}
}

\def\figXIV{%
\begin{figure}
\includegraphics*[height=0.35\textwidth]{currdetpp}
\caption{The profile of $j_\ch$ and $j_\sp$ on the intersection of 
  Py-detector/Cu interface and the xz' cut (defined on Fig.~\ref{f:cut}) for
  parallel magnetization state. The rest as
  in Fig.~\ref{f:currinj}.}
\label{f:currdet}
\end{figure}
}

\begin{document}

\title
{
Current distribution inside Py/Cu lateral spin-valve device
}

\author{J. Hamrle}
\affiliation{
FRS, The Institute of Physical and Chemical Research (RIKEN), 
2-1 Hirosawa, Wako, Saitama 351-0198, Japan}
\affiliation{CREST, Japan Science \& Technology Corporation, Japan}

\author{T. Kimura}
\affiliation{
FRS, The Institute of Physical and Chemical Research (RIKEN), 
2-1 Hirosawa, Wako, Saitama 351-0198, Japan}
\affiliation{CREST, Japan Science \& Technology Corporation, Japan}

\author{Y. Otani}
\affiliation{
FRS, The Institute of Physical and Chemical Research (RIKEN), 
2-1 Hirosawa, Wako, Saitama 351-0198, Japan}
\affiliation{CREST, Japan Science \& Technology Corporation, Japan}
\affiliation{ISSP, University of Tokyo, Kashiwa-shi, Chiba 277-8581, Japan}

\author{K. Tsukagoshi}
\affiliation{RIKEN, 2-1 Hirosawa, Wako, Saitama 351-0198, Japan}
\author{Y. Aoyagi}
\affiliation{RIKEN, 2-1 Hirosawa, Wako, Saitama 351-0198, Japan}

\date{\today}
\begin{abstract}
  We have investigated experimentally the non-local voltage signal
  (NLVS) in the lateral permalloy (Py)/Cu/Py spin valve devices with
  different width of Cu stripes. We found that NLVS strongly depends
  on the distribution of the spin-polarized current inside Cu strip in
  the vicinity of the Py-detector. To explain these data we have
  developed a~diffusion model describing spatial (3D) distribution of
  the spin-polarized current in the device. The results of our
  calculations show that NLVS is decreased by factor of 10 due to spin
  flip-scattering occurring at Py/Cu interface. The interface
  resistivity on Py/Cu interface is also present, but its contribution
  to reduction of NLVS is minor. We also found that most of the
  spin-polarized current is injected within the region 30\,nm from
  Py-injector/Cu interface. In the area at Py-detector/Cu interface,
  the spin-polarized current is found to flow mainly close on the
  injector side, with $1/e$ exponential decay in the magnitude within
  the distance 80\,nm.

\vspace*{1cm}

\end{abstract}

\maketitle

\section{Introduction}

Spintronics is a quickly evolving field providing the possibility to
manipulate spin degrees of freedom in the solid state systems
\cite{ziese,zut04} . Spin injection, transport and detection in metals
and semiconductors are of particular importance to construct effective
spintronic devices such as a spin battery \cite{bra02} and spin
torque transistor \cite{bau03} etc. Such devices have great advantages
over the conventional electronic devices because of additional spin
functionalities. To realize the device it is a key to obtain both
large spin-polarized current and spin accumulation. It is also
important to understand the diffusive nature of the spin-polarized currents in
multiterminal devices. 

Recently, the non-local probing method was proposed by Jedema {\it et
  al.} \cite{jed99, jed01, jed02} to extract only spin-polarized current
contribution from the spin-dependent phenomena and to reduce spurious
effects such as Hall effect and anisotropic magnetoresistance. They
succeeded in detecting the clear spin-accumulation signal in the
vicinity of the non(N)-/ferro(F)-magnetic planar junction by the
non-local spin-voltage (NLVS) even at room temperature \cite{jed01,
  jed02}. Furthermore, non-local technique maybe useful to induce
spin-injection magnetization reversal \cite{kat00} without the charge
current, leading to the solution for the energy dissipation problem
due to Joule heat.

Here, we study experimentally the distribution of the spin-polarized
current in non-local configuration.  So far, the spin-polarized
current transport is analytically investigated using one dimensional
(1D) Boltzmann diffusion model \cite{son87, val93, jed01, tak02,
  tak03}.  As these models predict too large NLVS, we have developed
formalism to calculate spatial (3D) distribution of spin-polarized
current.  However, large decrease of NLVS can not be attributed to
spatial distribution of spin-polarized current, and we attribute it to
spin scattering at Py/Cu interface.

\section{Device preparation and experimental results}
\label{s:sample}

\figI

We fabricated lateral spin-valve devices consisting of two Py wires
bridged by Cu strip by means of nano-fabrication techniques.
Figure~\ref{f:elimage} shows a scanning-electron-microscope (SEM)
image for one of the fabricated devices.  First, we fabricated both Py
wires of width $w_\py=120$\,nm and of thickness $t_\py=20$\,nm with
the spacing of $l_\cu=170$\,nm by electron-beam lithography and
lift-off technique.  Py layer was evaporated by an electron-beam gun
at 2 $\times 10^{-8}$ Torr.  Ends of the first Py wire are connected
to large pads pattern for assisting the nucleation of the domain wall,
although ends of the second one are flat-end shaped.  Hence, each Py
wire has different switching field.  

\figII
\figIII

Both Py wires are bridged by Cu strip of thickness $t_\cu=80$\,nm,
having widths $w_\cu=$100, 300 and 500\,nm for three different
devices.  Prior to Cu deposition, the Py surface was cleaned by Ar$^+$
bombardment and then sample was shortly taken out-of-vacuum to change
vacuum chamber. Then Cu was evaporated by resistance heating.  The
contact resistance of the interface was found ohmic and very low
indicating a transparent contact.  The conductivity of Cu is
$\sigma_{\cu,\mathrm{RT}}=48.1\times10^6\,\Omega^{-1}$m$^{-1}$,
$\sigma_{\cu,\mathrm{4K}}=131\times10^6\,\Omega^{-1}$m$^{-1}$ at room
temperature, 4 K, respectively.

Notice that the present Cu strip has smaller residual resistivity than
that of Jedema \textit{at el.} \cite{jed01}. The NLVS measurements were
performed at room temperature with the magnetic field applied parallel
along the Py-wires.

NLVS measurements were performed using a standard current-bias lock-in
technique at room temperature. We measured the NLVS as a function of
external magnetic field by using two different probe configurations,
called ``half'' and ``cross''.
The difference between both configurations is whether the current and
voltage probes are located on the same side or not as sketched in
Figs.~\ref{f:loc}(a) or \ref{f:loops}(a). The one-dimensional (1D)
diffusion model \cite{jed03,tak03} predicts that the obtained NLVS
should be the same.  However, as the spin-polarized current has the
spatial distribution, the NLVS shows the difference between both probe
configurations.

Figures~\ref{f:loops}(c) and \ref{f:loops}(e) show a NLVS 
for $w_\cu=100$\,nm with ``half'' and ``cross'' probe
configuration, respectively. The obtained difference of NLVS between
parallel and antiparallel magnetization (DNLVS) is 0.7\,m$\Omega$ and
0.6\,m$\Omega$ at room temperature, respectively.  Figures
\ref{f:loops}(d) and \ref{f:loops}(f) show NLVS for
$w_\cu=500$\,nm with ``half'' and ``cross'' configurations, providing
0.6 and 0.1\,m$\Omega$, respectively. 

Experimental values of DNLVS as a function of $w_\cu$ are presented on
Fig.~\ref{f:wcu}(a). Experimental data show that the difference
between ``cross' and ``half'' in DNLVS increases with increasing
$w_\cu$.

\figIV

The other parameters (at room temperature) used in our
calculations are as follow: Py conductivity
$\sigma_\py=7.3\times10^{6}\,\Omega^{-1}$m$^{-1}$, Py bulk spin
asymmetry coefficient $\beta=0.7$ ($\sigma_{\up,\py}=\sigma_\py
(1+\beta)/2$, $\sigma_{\dw,\py}=\sigma_\py(1-\beta)/2$)
(Refs.~\cite{dub99,ste97,hol98}), spin-flip lengths
$\lambda_\py=4.3$\,nm (Ref.~\cite{dub99}) and $\lambda_\cu=350$\,nm
(Ref.~\cite{jed03}).  Py wires have width $w_\py=120$\,nm, thickness
$t_\py=20$\,nm and separated by distance $l_\cu=170$\,nm.  The Cu
strip has thickness $t_\cu=80$\,nm with widths $w_\cu={100,\ 300,\ 
  500}$\,nm.

\section{1D calculations of DNLVS}
\label{s:curr1d}

In the literature, there are two models describing NLVS (and DNLVS)
inside metallic lateral spin-valve device: one given by Jedema \textit{et.\ 
al.}~\cite{jed03} and the other by Takahashi and Maekawa \cite{tak03}.
Both models approximate the device into 1D wire circuit, in which
the propagation of electrochemical potential $\mu_{\up/\dw}$ and
spin-polarized current $J_{\up/\dw}$ is described by standard
Valet-Fert model \cite{val93}.  At an intersection point of several
wires (hereafter called node), e.g.\ intersection of Cu and Py wires, the
boundary conditions, expressed as generalized Kirchoff's laws are
\begin{equation}
\begin{array}{rl}
\displaystyle
\sum_n J_{n,\up/\dw}&
\displaystyle
=0 \\
\displaystyle
\mu_{n,\up/\dw}&
\displaystyle
=\mathrm{const}_{\up/\dw}
\end{array}
\label{eq:kir}
\end{equation}
where $n$ is an index of all the wires connected to a~given node.
Hence, the $\mu_{\up/\dw}$ (which can be understand as a~voltage here)
is the same for each 1D wire attached to a given node, and
$J_{\up/\dw}$ is conserved while flowing through each node.

\figV

The model of Jedema \textit{et al.} \cite{jed03} has two assumptions,
which are not fulfilled in our case of Py/Cu device: (a) they assume
cross-sectional areas of all the wires in the device were the same
(i.e.\ they considered more continuity of up and down current
densities $j_{\up/\dw}$ than up and down currents $J_{\up/\dw}$ at
each node) and (b) they assume $(\lambda_F,\lambda_N) \gg (w_F,w_N)$,
where $(w_F,w_N)$ are widths of F, N wires, respectively.  The
comparison of DNLVS obtained from this model (when extended to the
case for different cross-sectional areas of wires) with our
experimental data is shown on Fig.~\ref{f:wcu}(b) (dashed-dot line),
showing that this model predicts about 40$\times$ large value than
experimental one.

These drawbacks were partly overcome by Takahashi and Maekawa
\cite{tak03}, assuming that (a) $\lambda_F\ll (w_F, w_N) \ll
\lambda_N$ and (b) that current at F/N interface is
homogeneous. Later we will show that assumption (b) is not correct for
ohmic junctions, but is correct for tunnel junctions.  Although they
derived their model from basic equations, the same results can be
obtained when both F-injector and F-detector, attached to N wire, are
described by a~standard 1D model, where F-wires have effective
cross-section area as of Py/Cu interface, i.e.\ in our case
$\tilde{S}_F=w_F w_N$.  The DNLVS calculated from this model is
presented on Fig.~\ref{f:wcu}(b) for the case with
interface resistance $R^\star_\mathrm{Py/Cu}=0$ (solid line) and
$R^\star_\mathrm{Py/Cu}=0.5$\,f$\Omega$m$^2$, $\gamma=0.7$
(Ref.~\cite{ste97}) (dashed line). Note that
$R^\up_{\mathrm{Py/Cu}}=2AR^\star_\mathrm{Py/Cu} (1-\gamma)$,
$R^\dw_{\mathrm{Py/Cu}}=2AR^\star_\mathrm{Py/Cu} (1+\gamma)$. This 1D
model describes quite well the experimentally observed DNLVS but
gives about 10~times larger magnitude than the experimental
results.

\section{3D calculation of spin-polarized current and electrochemical
potential}
\label{s:3dcalc}

\figVI

In order to understand the spin-polarized currents inside device in
detail, we have developed model calculating 3D distribution of
$\mu_{\up/\dw}$ and spin-polarized current density $j_{\up/\dw}$
inside the device \cite{ham04}. Our model is based on the 3D
electrical network of spin-dependent-resistance-elements (SDRE)
(Fig.~\ref{f:net3d}). The response of each SDRE is determined by 1D
models \cite{val93,jed03}.  As sketched in inset of
Fig.~\ref{f:net3d}, each SDRE consists of resistance for spin-up,
spin-down channels and spin-flip resistance shunting up and down
channels. This shunting resistance can be regarded as the
``probability'' that electron spins are flipped when passing SDRE.
A~boundary conditions at each node connecting SDRE are given by
Eq.~(\ref{eq:kir}).

In this
model we can also account surface or interface resistance
(scattering), $AR_\ss$ or $AR_\is$, respectively, shortcutting up
and down channels at the surface or interface.  For detailed
description of the formalism please see \cite{ham04}.

\subsection{Plausibility of 3D calculations}
\label{s:plau}

To estimate precision of our 3D calculations, we have calculated
$j_\updw$, $\mu_\updw$ and magnetoresistivity ratio (MR) in
Cu/Py(20)/Cu(20)/Py(20)/Cu \textit{multilayer} structure (dimensions in nm)
using different grid sizes for Py. The results of these calculations
should be identical with 1D Valet-Fert model \cite{val93}. We
investigate calculation precision only with grid size of Py, as
$\lambda_\py\ll\lambda_\cu$.

Figure~\ref{f:jr13} shows profile of spin-polarized current
$j_\sp=j_\up-j_\dw$ through antiparallel Py/Cu multilayer structure.
Used lateral grid size (i.e.\ grid distances parallel with Py/Cu
interfaces) is 10\,nm, perpendicular grid (i.e.\ grid perpendicular to
interfaces) is 1, 5, 10\,nm, giving $j_\sp$ precision inside Py being
4\%, 9\%, 15\%, respectively. 3D calculations gives larger value of MR
by 8\%, 16\%, 33\% than 1D calculation.  For in-plane grid size 5\,nm,
the MR is larger by 4\%, 11\%, 26\%. It shows that (i) with decrease
of grid size, $j_\sp$ and MR converge to correct values and
(ii) small perpendicular grid size is more important than in-plane one.

Figure~\ref{f:gridsize} shows a~dependence of DNLVS in the lateral
spin-valve structure on various lateral (i.e. parallel with substrate
surface) grid size. The simulated device is different than
real one; two Py wires of 15\,nm-thick and 50\,nm-width are separated
by a distance of 80\,nm and bridged by 55\,nm-thick, 50\,nm-width Cu strip. 

DNLVS has been calculated for perpendicular grid size 5\,nm
(square in Fig.~\ref{f:gridsize}) and 2.5\,nm (diamond), providing
larger DNLVS by 18\%, 11\%, respectively, with respect to the
converged DNLVS value.  In both cases, larger grid size leads to
larger DNLVS. 

\figVII

In all simulations of real structure, we used
perpendicular grid size 5\,nm, lateral grid size 10\,nm and in the
vicinity of Py/Cu interface lateral grid size 5\,nm. This grid
configuration is denoted by a~circle on the Fig.~\ref{f:gridsize},
providing agreement with DNLVS converged value 13\%, in agreement with
above discussion.  Unfortunately, in our calculations, we can not use
smaller grid size due to numerical limitations.  We conclude (i) precision
of our 3D calculations is about 20\% (ii) calculated DNLVS has tendancy
to be overestimated.

\subsection{3D calculations of DNLVS}
\label{s:nlvs}

\tabI

Figure~\ref{f:wcu}(b) presents DNLVS calculated from 3D model for
$AR^\star_{\py/\cu}=0$ (circle) and for
$AR^\star_{\py/\cu}=0.5$\,f$\Omega$m$^2$, $\gamma=0.7$
(Ref.~\cite{ste97}) (diamond). Both DNLVS have about the same shape
and slightly smaller magnitude compared to values from 1D model by
Takahashi and Maekawa \cite{tak03} (solid and dashed lines without
symbols).  In agreement with experiment, for larger $w_\cu$, the DNLVS
has different values in half and cross configuration, reflecting
inhomogeneous $j_\sp$ at position of the detector.  As will be shown in
Sec.~\ref{s:currflow}, the $j_\sp$ is also strongly inhomogeneous at
injector position. However, the approximative agreement between 3D
models and 1D models shows, that influence of inhomogeneous current
injection is not very important to magnitude of DNLVS.  As our 3D
models tends to overestimate DNLVS, we conclude that non-local
current injection decrease DNLVS, but only about 20\%. 

Both 1D and 3D models shows that the presence of $R^\star_{\py/\cu}$
together with large positive value of $\gamma$ increases DNLVS. When
$R^\star_{\py/\cu}>0$, $\gamma=0$, DNLVS decreases. It may be possible
that $\gamma>0$, but this contribution to DNLVS is smeared by other
contribution decreasing DNLVS. Therefore, in following we assume
$\gamma=0$. 

Now, let us discuss which mechanism decrease DNLVS.  To be more sure
with analysis, we take into account more experimental data (which are
going to be published elsewhere \cite{kim-unpub}) on two different
samples, fabricated exactly by a way as previous sample.
\begin{itemize}
\item %
  3-wires system consisting of two Py wires of 20\,nm-thick,
  100\,nm-width, separated by a distance of 400\,nm and bridged by
  80\,nm-thick 100\,nm-width Cu strip.  Between both Py wires, there
  is third 100\,nm-width wire [Fig.~\ref{f:loc}(b)], consisting either
  of Cu (having thickness 80\,nm), or Py wire (having thickness
  20\,nm), or there is no third-wire.
\item %
  system consisting of two Py 20\,nm-thick wires with different widths
  (200\,nm-width of injector and 100\,nm-width of detector), separated
  by 200\,nm and bridged by 250\,nm-width and 80\,nm-thick Cu strip.
  In this device, we measured both DNLVS and difference of
  \textit{local} voltage signal between parallel and antiparallel
  state (DLVS) [Fig.~\ref{f:loc}(c)]. In DLVS case, charge current
  flows through both Py wires.
\end{itemize}

In the following, we will discuss possible contributions coming from
(i) surface scattering on Cu, $AR_{\ss,\cu}$ (ii) surface scattering at
Py/Cu interface $AR_{\is,\py/\cu}$ (iii) interface non-polarized
resistance, $AR^\star_{\py/\cu}$.  The possible magnitude of each
contribution has been determined to fit DNLVS for $w_\cu=100$\,nm and
then compared with other experimental data.  All experimental data and
calculated values are summarized in Tab.~\ref{t:tab1}.

\subsubsection{Surface scattering on Cu}
Surface scattering on Cu is introduced by an resistance
  $AR_{\ss,\cu}$ shortcutting up and down channel on the Cu surface.
  To decrease DNLVS for $w_\cu=100$\,nm to experimental value,
  0.7\,m$\Omega$, Cu surface scattering has to be
  $R_{\ss,\cu}=0.15$\,f$\Omega$m$^2$ (when surface scattering is
  assumed on both side sides and top and bottom surface of Cu wire) or
  $R_{\ss,\cu,\side}=0.065$\,f$\Omega$m$^2$ (when surface scattering
  is assumed to be only on both sides of Cu). However, using those surface
  scattering resistances, the DNLVS calculated for 3-wires system
  (Tab.~\ref{t:tab1}) are too small compared with experiment, showing
  that this contribution is not a dominant one.

\subsubsection{Interface scattering on Py/Cu interface}
The properties of Py/Cu interface is here
  described by a presence of the interface layer, which has its own
  thickness $t_I$, spin-flip-length $\lambda_I$ and conductivity
  $\sigma_I$, spin-polarization $\gamma_I$ \cite{par00}. However, the
  interface properties should not depend on $t_I$ (this value is given
  \textit{ad-hoc} and is assumed as 1\,nm in our calculations).
  Therefore, it is profitable to express interface properties by
  $\delta_I=t_I/\lambda_I$ and $AR^\star_{\py/\cu}=t_I/\sigma_I$,
  which are independent on $t_I$ \cite{par00}. Physical meaning of
  $AR^\star$ is clear: $2R^\star(1-\gamma_I)$, $2R^\star(1+\gamma_I)$
  is a resistance of channel up, down through interface layer,
  respectively.  As physical meaning of $\delta_I$ is not so clear, we
  prefer to describe spin-flip scattering by interface scattering
  resistivity \cite{ham04}
  \begin{equation}
  \label{eq:ARs}
  AR_s = AR^\star\frac{4}{\delta\sinh\delta},
  \end{equation} 
  which means a~resistance shortcutting up and down
  channels on the interface.
  
  To decrease DNLVS to experimental value at $w_\cu=100$\,nm,
  different pairs of $AR^\star_{\py/\cu}$, $AR_{s,\py/\cu}$ can be used,
  as shown in Table~\ref{t:tab1}. When there is no interface
  resistance ($AR_{\py/\cu}^\star=0$), then
  $AR_{\is,\py/\cu}=2.6$\,f$\Omega$m$^2$. On the other hand, when
  $AR_{\is,\py/\cu}=\inf$ then $AR^\star_{\py/\cu}=15$\,f$\Omega$m$^2$.
  Both  $AR_{s,\py/\cu}$ and $AR^\star_{\py/\cu}$ contribute to
  decrease of DNLVS. 

  Table~\ref{t:tab1} and Figure~\ref{f:wcu}(a) shows that none
  combination of pairs $AR^\star_{\py/\cu}$, $R_{\is,\py/\cu}$
  describes perfectly all experimental values, however the agreement
  with all experimental data is within factor of 2-3. 
  Figure~\ref{f:wcu}(a) shows that with increasing value of
  $AR^\star_{\py/\cu}$, the difference between ``half'' and ``cross''
  DNLVS is reducing, reflecting more homogeneous injection of $j_\sp$
  over Py-inj/Cu interface. 
  
  The most relevant interface properties is a pair of values
  $AR^\star_{\py/\cu}=1$\,f$\Omega$m$^2$,
  $R_{\is,\py/\cu}=3.8$\,f$\Omega$m$^2$ ($\delta_{\py/\cu}=0.95$) as
  for this pair the mutual ratio between DNLVS's for 3-wire system
  (when middle wire is Cu, Py and nothing) agrees with experiment.
  Then all calculated values for 3-wires systems are about 1.8$\times$
  larger then experimental one. The disagreement by factor 1.8$\times$
  can be related to smaller value of $\lambda_\cu$ than expected
  350\,nm.  The 3-wire configuration with middle Py wire is
  particularly sensitive to $AR^\star_{\py/\cu}$, as its value
  determines, how large amount of $j_\sp$ is absorbed by the middle Py
  wire.
  
  Table~\ref{t:tab1} shows that experimental value of DNLVS at
  $w_\cu=300$\,nm is larger than calculated one, particularly for cross
  configuration (experimental DNLVS$_\mathrm{cross}=$0.3\,m$\Omega$,
  but calculated 0.14\,m$\Omega$). In another words, DNLVS($w_\cu$)
  decreases slower for experiment than for calculated value.  It is
  probably due to presence of a charge current $j_\ch$ at a~position
  of Py-detector for wider $w_\cu$, as will be shown in
  Sec.~\ref{s:top}. Non-zero $j_\ch$ inside detector probably causes
  some additive contribution to DNLVS, either due to AMR, either due
  to the scattering related with currents in-plane (CIP), i.e.
  currents flowing parallel with Py/Cu interface
  
  Figure~\ref{f:wcu}(b) also contain a dependence DNLVS($w_\cu$)
  calculated from extended model of Takahashi and Maekawa 
  for
  $AR^\star_{\py/\cu}=1$\,f$\Omega$m$^2$,
  $R_{\is,\py/\cu}=3.8$\,f$\Omega$m$^2$. We can see that there is
  a~good agreement with 3D calculations. It shows when $j_\sp$ is
  homogeneous on detector position, this model predicts a correct
  value of DNLVS.
  
  The last part of Table~\ref{t:tab1} shows an agreement between
  experimental and calculated values of DNLVS and DLVS, determined for
  $w_\cu=250$\,nm. We can see that for
  $AR^\star_{\py/\cu}=1$\,f$\Omega$m$^2$,
  $R_{\is,\py/\cu}=3.8$\,f$\Omega$m$^2$, all experimental values are
  about twice large compared to calculated one. Probably, here play
  role similar effects as discussed for DNLVS for $w_\cu=300$\,nm, as
  in this case calculated DNLVS is also twice smaller than
  experimental one.
  
  Resistance $AR^\star_{\py/\cu}=1$f$\Omega$m$^2$ is equal to
  resistance of 48\,nm of Cu or 7.3\,nm of Py. Furthermore, interface
  scattering $AR_{s,\py/\cu}=3.8$f$\Omega$m$^2$ corresponds to
  scattering by Cu at length 950\,nm and at length 2.5\,nm inside
  Py [Eq.~(\ref{eq:ARs})]. Especially second value shows that
  interface scattering is not so large, however, it is enough to
  decrease DNLVS by one order of magnitude.

In conclusion of this Section, we have shown that major contribution
to small DNLVS is due to interface scattering resistance $AR_{\is,\py/\cu}$,
shortcutting up and down channels at  Py/Cu interfaces. The
interface resistivity $AR^\star_{\py/\cu}$ is also presented, but its
contribution to decrease of DNLVS is only minor one.  Such a~large
interface spin-scattering has not been observed in \cite{dub99,
  ste97}. It can be related with two factors:
\begin{itemize}
\item[1.] Quality of our Py/Cu interface is lower than in \cite{dub99,
    ste97}. In our fabrication process, there are two steps which
  could decrease interface quality. On top of Py we deposited and
  removed photoresist to pattern Cu wire. Before Cu deposition, the
  surface was cleaned by Ar$^+$ bombardment and then device shortly
  taken out of vacuum to change vacuum chamber. Notice, that Jedema et
  al.~\cite{jed03} has used very similar fabrication process as we
  did.
  
\item [2.] Contribution of Py/Cu interface spin scattering is missing
  in previous works, investigating Py/Cu system by means of
  magnetoresistivity ratio (MR) \cite{dub99, ste97}. Note that MR is
  sensitive to value of $j_\sp$ passing free layer rather that to
  value of spin accumulation $\Delta\mu$ at position of free layer
  \cite{ham04}. As we have shown \cite{ham04}, system can provide
  large MR (when large $j_\sp$ flow through free layer) although
  $\Delta\mu$ at position of free layer can vanish. When $\Delta\mu$
  vanishes, then shortcutting of up and down channels takes no effect
  and so interface spin-scattering does not occurs at the interface.
  In such a case, the MR (up to some limit) is insensitive to
  spin-scattering on the free-layer/non-magnetic-layer interface.
  
  On the other hand, non-local technique is particularly sensitive to
  $\Delta\mu$ at detector/non-magnetic-metal interface. When interface
  spin-scattering is presented in this case, it significantly reduces
  DNLVS.
  
  Hence, it may be possible, that small interface scattering is
  presented in both MR and non-local measurements, but did not take a
  place in case of MR measurements.
\end{itemize}

\figVIII
\figIX
\figX
\figXI
\figXII
\figXIII
\figXIV

\section{Current flows inside lateral spin-valve structure}
\label{s:currflow}

In this Section, we present in detail the current inhomogeneity inside
lateral spin-valve structure.  Figure~\ref{f:cut} shows a~sketch of
the device with indicated cut planes, on which the calculated current
densities are presented on Figures~\ref{f:frontinj}, \ref{f:topcu},
\ref{f:wcu300}, \ref{f:sidecu} and discussed in following Sections
\ref{s:inj}--\ref{s:det}. The presented current densities were
calculated for parallel magnetizations and for our best interface
description $AR^\star_{\py/\cu}=1$\,f$\Omega$m$^2$,
$R_{\is,\py/\cu}=3.8$\,f$\Omega$m$^2$. For antiparallel
magnetizations, we get very similar current flows as for parallel one.
This is in agreement with 1D models of non-local devices
\cite{jed03,tak03}, where current flows are exactly the same for
parallel and antiparallel magnetic states.

\subsection{Current description near P\lowercase{y}-injector}
\label{s:inj}

Figure~\ref{f:frontinj} shows current density on the yz cut, which is
taken in the center of the Py-injector wire (Fig.~\ref{f:cut}). Cuts
(a--d) correspond to the cases for up and down current densities
$j_\up$, $j_\dw$, respectively, for charge current density
$j_\ch=j_\up+j_\dw$ and for spin-polarized current density
$j_\sp=j_\up-j_\dw$.  All cuts show that the current is
injected rather sharply through Py-injector/Cu interface and then
quickly spreads into the whole volume of Cu wire.

The values of $j_\ch$ and $j_\sp$ at the intersection of yz cut and
Py-injector/Cu interface are presented on Fig.~\ref{f:currinj}. The
profile is shown for the device with $AR^\star_{\py/\cu}=0$,
$AR_{s,\py/\cu}=\inf$ (circles) and for
$AR^\star_{\py/\cu}=1$\,f$\Omega$m$^2$,
$R_{\is,\py/\cu}=3.8$\,f$\Omega$m$^2$ (triangles and diamond).  Due to
spin-flip-scattering on the interface, the $j_\sp$ flowing to the
interface from Py side (triangles-down) is about twice larger than
from $j_\sp$ outgoing the interface at Cu side (triangles-up).

It is shown that both $j_\ch$ (open symbols) and $j_\sp$ (solid
symbols) are sharply injected within the distance of 25\,nm, 35\,nm
from the Py/Cu edge for $AR^\star_{\py/\cu}=1$\,f$\Omega$m$^2$,
$AR_{s,\py/\cu}=3.8$\,f$\Omega$m$^2$ and $AR^\star_{\py/\cu}=0$,
$AR_{s,\py/\cu}=\inf$ respectively. This different
'length-of-injection' is only due to different values of
$AR^\star_{\py/\cu}$, and is nearly independent on $AR_{s,\py/\cu}$.
When $R^\star_{\py/\cu}$ is large then obviously the current is more
spread over the interface and for tunnel contacts is can be considered
as homogeneous.  Furthermore, $j_\sp$ is positive only in the distance
of 25\,nm or 35\,nm from the Py-injector/Cu edge, and then its value
becomes negative. This means that in this region the injector
reabsorbs a~small part of the injected spin-polarized current, which decreases
the spin-injection efficiency.

For different values of $w_\cu$, the 'length-of-injections' are very
similar to those presented in Fig.~\ref{f:currinj}. It should be
noticed that this sharp injection occurs in consequence of small Py
conductivity, $\sigma_\py\ll\sigma_\cu$ and small thickness of Py wire
$t_\py<(w_\cu,t_\cu)$. In other words, larger $t_\py$ increases
homogeneity of the injected current.

\subsection{Top view on C\lowercase{u}}
\label{s:top}

Figure~\ref{f:topcu} presents current density on xy cut (defined in
Fig.~\ref{f:cut}) taken at the depth of 12.5\,nm from top surface of
the Cu wire. As already discussed, current is sharply injected at
Cu/Py-injector edge and hence $j_\up$ and $j_\sp$ spread into the Cu
wire from this edge [Fig.~\ref{f:topcu}(a)(d)]. 

Figure~\ref{f:topcu}(d) also shows that for $w_\cu=100$\,nm, $j_\sp$
at the position of the detector is fairly uniform in the $y$-direction,
i.e.\ in the direction parallel to the Py wire. When $j_\up$ reaches
detector, it is successively spin-scattered due to very short
spin-diffusion length $\lambda_\py$ and then current flows
homogeneously back as $j_\dw$ [Fig.~\ref{f:topcu}(a)(b)].

Due to the sharp current injection, $j_\ch$ makes a whirl in the
'diffusive' part of the Cu wire, where no charge current was expected
[Figure~\ref{f:topcu}(c)]. In the present case ($w_\cu=100$\,nm), the
value of $j_\ch$ originating from this whirl at detector position is
negligible compared to $j_\sp$. However when $l_\cu< w_\cu$ ($l_\cu$
being distance between Py wires), then $j_\ch\gtrapprox j_\sp$ at
detector position.  This can be seen on Fig.~\ref{f:wcu300}(a) for
$w_\cu=300$\,nm. Fig.~\ref{f:wcu300}(b) shows that also $j_\sp$ for
$w_\cu=300$\,nm is inhomogeneous at detector position, having maximal value
at one side of Cu/Py-detector interface. This explains different
values between ``cross'' and ``half'' configurations.

\subsection{Current description near P\lowercase{y}-detector}
\label{s:det}

Figure~\ref{f:sidecu} presents current density on the xz cut (defined
in Fig.~\ref{f:cut}) which is taken 7.5\,nm from the side of Cu wire.
Figure~\ref{f:sidecu}(a) shows that the flow of $j_\up$ into detector
is also inhomogeneous and is dominant at the side of Py-detector/Cu,
which is close to the injector. As already mentioned above, due to
very short spin-diffusion length $\lambda_\py=4.3$\,nm, $j_\up$
flowing into the detector is immediately reversed inside Py-detector
and coming back as $j_\dw$ [Fig.~\ref{f:sidecu}(b)].  This can be
understand as a~resistance shunting (or ``short-cutting'') the up and
down channels. This also explains the behaviour of $j_\sp$
[Fig.~\ref{f:sidecu}(d)], whose flow is absorbed by the detector.

Figure~\ref{f:currdet} shows $j_\ch$ and $j_\sp$ on the
intersection between Py-detector/Cu interface and xz$'$ cut (defined
on Fig.~\ref{f:cut}), which is taken at the center of Cu wire. The
vertical dash-dot lines show the position of edges inside Py-detector
wire embedded in the Cu wire, i.e.\ ranges $x\in(-20,0)$ and
$x\in(120,140)$ corresponds to side part of Py wire, although range
$x\in(0,120)$ represents bottom part of Py-detector wire. We can see
that both $j_\sp$ and $j_\ch$ are inhomogeneous, decaying
approximately exponentially with $1/e$ decrease length $80$\,nm. This
decay is mainly result of the competition between Cu conductivity
$\sigma_\cu$ and spin-flip scattering inside Py and Py/Cu interface.

When the interface resistances $AR^\star_{\py/\cu}=1$\,f$\Omega$m$^2$,
$AR_{s,\py/\cu}=3.8$\,f$\Omega$m$^2$ are introduced, $j_\sp$ flowing
to the detector is decreased (and hence DNLVS is decreased), as can be
seen on Fig.~\ref{f:currdet}. Due to presence of $AR_{s,\py/\cu}$, the
current flowing to the interface from Cu side (triangle-up) is about
$3\times$ larger than one outgoing to the injector (triangle down).
Hence, $2/3$ of $j_\sp$ entering detector are shortcut, absorbed by
a interface.

There is also $j_\ch$ at Py-detector/Cu interface, having value
about 10\% of $j_\sp$. The $j_\ch$ originates because
$j_\up$ and $j_\dw$ are injected to/ejected from the Py-detector at
slightly different position, i.e.\ $j_\ch$ has negative value around
$x\gtrsim 0$ and positive at $x\lesssim 0$. It means that part of
$j_\up$ current, which is injected to Py-detector from the side of the
Py wire, is ejected as $j_\dw$ from its top part.

\section{Conclusion}

We have fabricated lateral spin-valve devices consisting of the
permalloy (Py) and Cu wires. We have observed that the difference of
the non-local voltage signal (DNLVS) between parallel and antiparallel
magnetization has different values for "half" and "cross"
configurations.  The difference between these two configurations
increases when the width of the Cu stripe increases.

To understand observed behaviour in detail, we have developed formalism
calculating spatial (3D) distribution of the spin-polarized current
and electrochemical potential inside the device. We found that the
current distribution inside lateral spin-valve device with
ohmic-contact is rather complex interplay between geometry and
electrical properties of all the involved materials.  

Despite of those large current inhomogeneities, the DNLVS calculated
from our 3D model are in a good agreement with 1D model given by
Takahashi et Maekawa \cite{tak03}. However, both 1D and 3D predicts
about 10$\times$ larger DNLVS than experimental values.  We have
attributed the smallness of DNLVS to interface scattering resistance
$AR_{s,\py/\cu}=3.8$f$\Omega$m$^2$ shortcutting up and down channels
at Py/Cu interface. On one hand, this value of $AR_{s,\py/\cu}$
decrease DNLVS by factor of 10. On the other hand, it corresponds only
to scattering which occurs inside Py on distance of 2.5\,nm.  When
this interface scattering resistance can be reduced, DNLVS may be
enhanced significantly. The fact, that such a interface resistivity
has not been observed before \cite{dub99,ste97} may be related either
to lower quality of our interface, either to insensitivity of MR to
small surface scattering in some cases.

Interface resistance $AR^\star_{\py/\cu}=1$f$\Omega$m$^2$ is also
presented at Py/Cu interface, but its contribution to smallness of
DNLVS is minor. The value of this resistance mainly modifies the
$j_\sp$ inhomogeneity in the structure. Using this description of 
Py/Cu interface, we found agreement with all our experimental data
(local and non-local voltage signals measured on systems with two or
three Py wires) within factor of two. 

The current is injected from Py-injector to Cu sharply, within the
distance of 30\,nm. Part of the injected spin-polarized current is
reabsorbed by injector itself.  Current flow over Py-detector/Cu
interface is also inhomogeneous, having the largest value on the side
of Py-detector close to injector and decaying approximately
exponentially with $1/e$ decrease within the distance $80$\, nm.

\bibliography{../spinbib}

\begin{thebibliography}{10}

\bibitem{ziese}
M.~Ziese and M.~Thornton, editors,
\newblock {\em Spin electronics} (Springer, Berlin, 2001).

\bibitem{zut04}
I.~Zutic, J.~Fabian, and S.~{Das Sarma},
\newblock Rev. Mod. Phys. {\bf 76}, 1 (2004).

\bibitem{bra02}
A.~Brataas, Y.~Tserkovnyak, G.~E.~W. Bauer, and B.~I. Halperin,
\newblock Phys. Rev. B {\bf 66}, 060404R (2002).

\bibitem{bau03}
G.~E.~W. Bauer, Y.~Tserkovnyak, D.~Huertas-Hernando, and A.~Brataas,
\newblock Phys. Rev. B {\bf 67}, 094421 (2003).

\bibitem{jed99}
F.~J. Jedema, B.~J. van Wees, B.~H. Hoving, A.~T. Filip, and T.~Klapwijk,
\newblock Phys. Rev. B {\bf 60}, 16549 (1999).

\bibitem{jed01}
F.~J. Jedema, A.~T. Filip, and B.~J. van Wees,
\newblock Nature {\bf 410}, 345 (2001).

\bibitem{jed02}
F.~J. Jedema, H.~B. Heersche, A.~T. Filip, J.~J.~A. Baselmans, and B.~J. van
  Wees,
\newblock Nature {\bf 416}, 713 (2002).

\bibitem{kat00}
J.~A. Katine, F.~J. Albert, R.~A. Buhrman, E.~B. Myers, and D.~C. Ralph,
\newblock Phys. Rev. Lett. {\bf 84}, 3149 (2000).

\bibitem{son87}
P.~C. van Son, H.~van Kempen, and P.~Wyder,
\newblock Phys. Rev. Lett. {\bf 58}, 2271 (1987).

\bibitem{val93}
T.~Valet and A.~Fert,
\newblock Phys. Rev. B {\bf 48}, 7099 (1993).

\bibitem{tak02}
S.~Takahashi and S.~Maekawa,
\newblock Phys. Rev. Lett. {\bf 88}, 116601 (2002).

\bibitem{tak03}
S.~Takahashi and S.~Maekawa,
\newblock Phys. Rev. B {\bf 67}, 052409 (2003).

\bibitem{jed03}
F.~J. Jedema, M.~S. Nijboer, A.~T. Filip, and B.~J. van Wees,
\newblock Phys. Rev. B {\bf 67}, 085319 (2003).

\bibitem{dub99}
S.~Dubois {\em et~al.},
\newblock Phys. Rev. B {\bf 60}, 477 (1999).

\bibitem{ste97}
S.~Steenwyk, S.~Hsu, R.~Loloee, J.~Bass, and W.~{Pratt, Jr.},
\newblock J. of Magn. Magn. Mater. {\bf 170}, L1 (1997).

\bibitem{hol98}
P.~Holody {\em et~al.},
\newblock Phys. Rev. B {\bf 58}, 12230 (1998).

\bibitem{ham04}
J.~Hamrle, T.~Kimura, T.~Yang, and Y.~Otani,
\newblock cond-mat/0409309.

\bibitem{kim-unpub}
T.~Kimura, J.~Hamrle, and Y.~Otani,
\newblock to be published.

\bibitem{par00}
W.~Park {\em et~al.},
\newblock Phys. Rev. B {\bf 62}, 1178 (2000).

\end{thebibliography}
\bibliographystyle{../h-physrev}

\end{document}